\documentclass[twocolumn,amsmath,amssymb]{revtex4-1}

\usepackage[utf8]{inputenc}
\usepackage{graphicx}
\usepackage{bm}
\usepackage{amsmath}
\begin{document}
\title{Attosecond Transient Absorption Spectroscopy without Inversion Symmetry}

\author{L. Drescher}
\email{lorenz.drescher@mbi-berlin.de}
\author{M.J.J. Vrakking}
\author{J. Mikosch}

\affiliation{Max-Born-Institut f\"ur nichtlineare Optik und Kurzzeitspektroskopie, Max-Born-Strasse 2A, 12489 Berlin, Germany}
\date{\today}

\begin{abstract}
Transient absorption is a very powerful observable in attosecond experiments on atoms, molecules and solids and is frequently used in experiments employing phase-locked few-cycle infrared and XUV laser pulses derived from high harmonic generation. We show numerically and analytically that in non-centrosymmetric systems, such as many polyatomic molecules, which-way interference enabled by the lack of parity conservation leads to new spectral absorption features, which directly reveal the laser electric field. The extension of Attosecond Transient Absorption Spectroscopy (ATAS) to such targets hence becomes sensitive to global and local inversion symmetry. We anticipate that ATAS will find new applications in non-centrosymmetric systems, in which the carrier-to-envelope phase of the infrared pulse becomes a relevant parameter and in which the orientation of the sample and the electronic symmetry of the molecule can be addressed.
\end{abstract}

\maketitle
\section{Introduction}
Attosecond Transient Absorption Spectroscopy (ATAS)~\cite{goulielmakis-2010,ramasesha2016:anrevphyschem} is emerging as one of the most potent techniques in attosecond science, since it takes advantage of both the appealing temporal and spectral properties of attosecond XUV pulses. ATAS has initially been applied to atoms and has shown its versatility in numerous studies~\cite{goulielmakis-2010,gruson-2016,ott-2013,kaldun2016:sci,chini-prl:2012,chini-2013,holler2011:prl}. It is very recently starting to make an impact also in molecular science \cite{warrick2016:jpca,warrick2017:cpl,drescher-2019,kobayashi2019:science,timmers2019:natcom}, as well as in studies of the condensed phase, where sub-cycle dependent modifications of the material dielectric function have been investigated~\cite{schultze-2013}, paving the way towards petahertz electronics~\cite{krausz-2014a}. As ATAS evolves beyond atomic systems, new aspects emerge that result from the anisotropic nature of more complex structures, as explored in recent theory papers \cite{baekoj2015:pra,baekhoj-2016,hollstein2017:pra,badanko-2018a,rorstad2018:pra}. For example, Badankó~\textit{et al.} \cite{badanko-2018a} investigated the importance of the orientation of the transition dipole moment in non-adiabatic molecular dynamics, and Rørstad~\textit{et al.} \cite{rorstad2018:pra} studied ATAS of polar molecules, discovering Light-Induced Structures (LIS) near bright rather than dark states and a ladder structure in the spectra that is spaced by the infrared (IR) photon energy.

The most relevant variables which control ATAS in investigations to date are: (i) the time delay $\tau$ between the attosecond pulse (or pulse train) and the IR field and (ii) the intensity of the IR field. In contrast, effects which are governed by the Carrier-to-Envelope Phase (CEP) of the IR pulse have not been reported. The CEP of a laser pulse is the phase between the carrier wave and the position of the intensity envelope and has become a routine (yet sophisticated) control parameter in attosecond experiments \cite{wirth2011:sci}. 

Here we identify features in numerically obtained ATAS spectrograms for a non-centrosymmetric model system, which oscillate as a function of XUV-IR time delay with a period 2$\pi$/$\omega_{\mathrm{IR}}$ ($\omega_{\mathrm{IR}}$-oscillations). This is in contrast with the oscillations with a period $\pi$/$\omega_{\mathrm{IR}}$ (2$\omega_{\mathrm{IR}}$-oscillations) that have thus far been seen in ATAS experiments on atoms and molecules. We explain the origin of these features in the spectrograms by extending a recently developed adiabatic model \cite{rorstad-2017} to non-centrosymmetric systems. The adiabatic model is then applied to dissect the ATAS spectrograms.

As we will show, the $\omega_{\mathrm{IR}}$-oscillations in the absorbance depend on a broken inversion symmetry, both locally (i.e. in a molecular fixed frame) and globally (i.e. in the laboratory frame). We hence anticipate new applications for ATAS in non-centrosymmetric systems: In oriented molecular samples, the CEP-dependent signals allow an all-optical diagnostic of the IR electric field and CEP-stability. On the other hand, with CEP-stable IR, the orientation of a (molecular) ensemble and the (evolving) electronic symmetry can be accessed.

In ATAS of atoms most of the observed phenomena can be understood by considering laser-induced couplings among a limited set of bound states \cite{wu-2016}. While the XUV laser field couples the ground state to one or more bright states, the IR laser field couples these to further states that are not directly accessible from the ground state via a dipole-allowed transition (dark states). The lack of coupling of the dark states to the ground state is ensured in atoms by parity selection rules. In atoms all states have a well-defined parity and the excitation of a particular state by both an even number of photons (e.g. the combination of an XUV photon and an IR photon) and an odd number of photons (e.g. an XUV photon only) is not possible. This is a direct consequence of the Laporte rule~\cite{wigner-1931}, which states that parity has to change in a dipole-allowed electronic transition. 

In molecules that lack centrosymmetry, parity can no longer be defined. In such systems two states can be coupled by both an odd and an even number of photons. In other words, excitation pathways then need to be taken into account where the XUV pulse coherently excites states that are then coupled by either an odd or an even number of IR photons. A three-level model system consisting of a ground state $E_0$ and two excited states $E_1$, $E_2$ captures most of the essential physics \cite{wu-2016}. It is described by the Hamiltonian:
\begin{equation}
\hat{H} = \left[\begin{matrix} E_0 & \vec d_{01}\cdot\vec\varepsilon_\textrm{XUV}(t) & \vec d_{02}\cdot\vec\varepsilon_\textrm{XUV}(t) \\ \vec d_{10}\cdot\vec\varepsilon_\textrm{XUV}(t) & E_1 & \Omega(t) \\ \vec d_{20}\cdot \vec\varepsilon_\textrm{XUV}(t) & \Omega^\ast(t) & E_2\\ \end{matrix}\right],\label{eq:full_hamiltonian}
\end{equation}
where
$\Omega(t) =\vec{d}_{12}\cdot\vec\varepsilon_{IR}(t)$,
$\vec d_{nm}=\vec d_{mn}^\ast$ are the transition dipole moments between the levels $n$ and $m$ and $\vec\varepsilon_{\textrm{XUV}}(t)$, $\vec\varepsilon_{\textrm{IR}}(t)$ are the time-dependent electric fields of the XUV and the IR pulse (see Appendix~\ref{ax:tdse}). While molecular transition dipole moments and electric fields generally are described by three dimensional vectors, we will first consider that the molecular frame (to which the dipole moment is fixed) is perfectly spatially oriented parallel to the electric fields in the laboratory frame. We then express them as scalars in our calculations by projection onto a common reference axis, i.e. $d_{nm} = \vec{d}_{nm}\cdot \vec{e}_z$ and $\varepsilon(t)=\vec{\varepsilon}(t)\cdot \vec{e}_z$.
The effect of this orientation against a more general alignment is discussed in section~\ref{sec:discussion}. Here and elsewhere, all equations are given in atomic units. 

In order to be able to contrast our results for non-centrosymmetric ATAS with the well-studied case of ATAS of the Helium atom, we choose parameters analogous to the latter, i.e. $E_0$ = 0\,eV $\sim$ He(1s$^2$), $E_1$ = 21.22\,eV $\sim$ He(1s2p), $E_2$ = 20.62\,eV $\sim$ He(1s2s), $d_{01}$ = 0.33\,a.u. and $d_{12}$ = 2.75\,a.u. 

The parameter $d_{02}$ depends on the symmetry of the model system, i.e. $d_{02} = 0$ for the centrosymmetric He atom and $d_{02} \neq 0$ for the non-centrosymmetric model system that we will consider here. In the latter case, we make the arbitrary choice to set $d_{02}=d_{01}=$0.33\,a.u. to ensure an equal population of both excited states by the XUV pulse.

\section{Numerical Solution}
We solve the time-dependent Schrödinger equation (TDSE) for the three-level problem described in Eq.~\eqref{eq:full_hamiltonian}:
\begin{equation}
    i\frac{\partial}{\partial t}\lvert\Psi(t)\rangle = \hat{H} \lvert\Psi(t)\rangle = \hat{H} \sum_{n=0}^2 c_n(t)e^{-iE_nt} \lvert\Phi_n\rangle.\label{eq:TDSE}
\end{equation}
Note that the presence of the phase-term $e^{-iE_nt}$ in this equation implies that $c_n(t)$ describes a slow variation of the amplitude of a given state due to coupling to other states. In the absence of such couplings, $c_n(t)$ is a constant. To account for the decay of excited states and finite spectral resolution, an imaginary term $i\frac{\Gamma}{2}$ is added to the excited states energy. While the He excited states lifetime is on the nanosecond scale, the lifetime is set to 30\,fs to visualize the delay-dependend regions of ATAS in a single spectogram~\cite{wu-2016}.

To obtain the time-dependent amplitudes $c_n(t)$, we follow the \textit{ansatz} of Eq.~\eqref{eq:TDSE} and get the system of ordinary differental equations (ODE):
\begin{widetext}
\begin{equation}
i\frac{\partial}{\partial t}
\left(
\begin{matrix} c_0 \\ c_1 \\ c_2
\end{matrix}
\right) = 
\left[
\begin{matrix}
 0 & d_{01}\varepsilon_\textrm{XUV}(t)e^{-iE_1t} & d_{02}\varepsilon_\textrm{XUV}(t)e^{-iE_2t} \\ 
 d_{10}\varepsilon_\textrm{XUV}(t)e^{iE_1t} & i\frac{\Gamma}{2} & \Omega(t)e^{i(E_1-E_2)t} \\ 
 d_{20}\varepsilon_\textrm{XUV}(t)e^{iE_2t} & \Omega(t)^\ast e^{i(E_2-E_1)t} & i\frac{\Gamma}{2}\\ 
 \end{matrix}
 \right]
 \left(
 \begin{matrix} 
 c_0 \\ c_1 \\ c_2
 \end{matrix}
 \right).\label{eq:interaction_ODE}
\end{equation}
\end{widetext}
The ODE is solved for a given XUV-IR time-delay by forward integrating in time using a Runge-Kutta-Dormand-Prince method of 5th order with adaptive step-size control.
For ease of notation, we assume a real-valued transition dipole moment for the rest of the discussion, i.e. $d_{nm} = d_{mn}$.

Knowing the full time-evolution of the system, the time-dependent dipole moment can be calculated:
\begin{equation}
    d(t) = \langle \Psi(t)\vert d\vert\Psi(t)\rangle = \sum_{n,m}c_n^\ast(t)c_m(t) d_{nm}e^{i(E_n-E_m)t}.\label{eq:time-dipole}
\end{equation}

The spectral representation of the time-dependent dipole can be calculated by a Fourier transformation:
\begin{equation}
    d(\omega) = \frac{1}{2\pi} \int_{-\infty}^\infty d(t) e^{i\omega t}\,dt = \mathcal{F}[d(t)](\omega).
\end{equation}

Since $d(t)$ is a real quantity, the spectral representation obtained via Fourier transformation is Hermitian, i.e. $d(-\omega) = d^\ast(\omega)$, meaning that the full spectral information is contained at either positive or negative frequencies.

The spectral response, i.e. the absorption or emission probability per unit frequency $\omega$ of a single molecule, is obtained by Fourier transformation of the temporal evolution of the dipole moment and the electric field according to~\cite{wu-2016}
\begin{equation}
    S(\omega,\tau) = \operatorname{Im}\left[\frac{\mathcal{F}[d(t)](\omega)}{\mathcal{F}[\varepsilon_\textrm{XUV}(t)](\omega)}\right].\label{eq:observable}
\end{equation}

Of interest in ATAS is the change of the spectral response, $\Delta S(\omega,\tau)$, i.e. the difference between the delay-dependent two-color response and the static XUV-only response. 

When the calculations are performed setting $d_{02}$ = 0, $\Delta A(\omega,\tau)$ displays a number of oscillatory features (figure\,\ref{fig:1}(a1)) that are well-established in the literature and that are defined by a $2\omega_{\mathrm{IR}}$ frequency that indicates the dominant role of processes involving two IR photons. In contrast, when choosing $d_{02} \neq 0$ (non-centrosymmetric case) the obtained spectrogram is dominated by features that oscillate as a function of $\tau$ with the periodicity of the IR field (figure\,\ref{fig:1}(b1)). They are discussed in detail below. The $2\omega_{\mathrm{IR}}$ features observed for the centrosymmetric case (figure\,\ref{fig:1}(a1)) are still present, as seen upon closer inspection. 

Further insight can be obtained by varying the IR field strength and the transition dipole moment $d_{12}$ in the simulation, i.e. $\Omega(t)$. While the modulation amplitudes of the $2\omega_{\mathrm{IR}}$ components depend quadratically on $\Omega(t)$, the amplitudes of the $\omega_{\mathrm{IR}}$ component depend only linearly on $\Omega(t)$. This explains the predominance of the non-centrosymmetric features in figure\,\ref{fig:1}(b1): Since for the chosen transition dipole moment and IR field strength $\Omega(t)\ll 1$, linearly dependent effects are much stronger than effects that depend quadratically on $\Omega(t)$.

\section{Adiabatic Solution}

For further insight, the TDSE for the three-level system is solved analytically using the adiabatic basis
\begin{equation}
\vert \Psi(t) \rangle = b_0(t)\vert \varphi_0 \rangle + c_+(t) e^{i\theta_{+}}\vert\varphi_+(t)\rangle + c_-(t) e^{i\theta_{-}}\vert\varphi_-(t)\rangle,\label{eq:adiabatic_basis}
\end{equation}
where $\vert\varphi_{\pm}(t)\rangle$ denote adiabatic excited eigenstates with time-dependent eigenenergies $E_{\pm}$, which are obtained by diagonalizing a reduced two-level Hamiltonian including only the excited states (1 and 2), justified since the IR-induced dynamics only involve these two XUV-excited states. In Eq.~\eqref{eq:adiabatic_basis}, the dynamic phase is given by $\theta_{\pm} = -\int_{\tau}^{t}dt'E_{\pm}(t')$. This basis allows to treat the XUV excitation perturbatively, while considering an adiabatic evolution of the IR-induced dynamics. Also here the CEP of the IR field is set to zero and perfectly oriented molecules are discussed first. We define a state mixing angle $\alpha(t)$, given by  
\begin{equation}
    \tan\alpha(t) = \frac{\Omega(t)}{\Delta+\sqrt{\Delta^2+\Omega(t)^2}} \approx \frac{\Omega(t)}{2\Delta+\Omega(t)^2/(2\Delta)},\label{eq:mixing_angle}
\end{equation}
with $\Delta = (E_1-E_2)/2$. The time- and intensity-dependent mixing-angle $\alpha(t)$ defines the projection of the time-dependent adiabatic states onto the time-independent field-free states:
\begin{equation}
\begin{aligned}\vert\varphi_+(t)\rangle&=\hphantom{-}\cos\alpha(t)\vert\Phi_1\rangle+\sin\alpha(t)\vert\Phi_2\rangle,\\
\vert\varphi_-(t)\rangle&=-\sin\alpha(t)\vert\Phi_1\rangle+\cos\alpha(t)\vert\Phi_2\rangle.\label{eq:adiabatic_eigenstates}
\end{aligned}
\end{equation}

For $\Omega(t) \to 0$ the mixing angle $\alpha(t) \to 0$ and $\vert\varphi_+\rangle$ and $\vert\varphi_-\rangle$ become the field-free excited states $\vert\Phi_1\rangle$ and $\vert\Phi_2\rangle$, respectively.

To find the coefficients $c_+(t)$, $c_-(t)$ within the adiabatic basis, the wavefunction Eq.~\eqref{eq:adiabatic_basis} is expanded in the field-free basis Eq.~\eqref{eq:adiabatic_eigenstates} and inserted into the TDSE (see Appendix~\ref{cha:theory_adiabatic} for further details). A perturbative approach to the XUV excitation is used to solve for the coefficients. Within a sudden approximation for the XUV excitation \cite{wu-2016,rorstad-2017}, i.e. $\varepsilon_\textrm{XUV}(t) = \delta(t-\tau)$, the time-dependent dipole moment can be written in a compact form using an index notation:
\begin{equation}
d(t) = i \vartheta(t-\tau)e^{-\frac\Gamma2 (t-\tau)}\sum_{n\neq m}^{1,2} e^{(-1)^n\,i\,\varphi(t,\tau)} \bigg[ C_n + N_n \bigg]+\textrm{c.c.}, \label{eq:adia_dipole}
\end{equation}
with
\begin{subequations}
\begin{align}
C_n \, &=& -d_{0n}^{2} \, e^{iE_n(t-\tau)} \, \cos{\alpha(\tau)}\cos{\alpha(t) } \hspace{1.5cm} \nonumber \\ && \hspace{0cm} -  d_{0n}^2 \, e^{iE_m(t-\tau)} \, \sin{\alpha(\tau) } \sin{\alpha(t) }, \label{eq:Sn} \\ \nonumber \\
N_n \, &=& (-1)^n \, d_{01} d_{02} \, e^{iE_n(t-\tau)} \, \bigg( \sin{\alpha(\tau) } \cos{\alpha(t) } \nonumber \\ && \hspace{1cm} + \cos{\alpha(\tau) } \sin{\alpha(t) } \bigg), \label{eq:An}
\end{align}
\nonumber
\end{subequations}
where $\textrm{c.c.}$ denotes the complex conjugate and $\vartheta(t-\tau)$ is the Heaviside function, which represents the step-like excitation by the XUV pulse, $\Gamma$ is the finite lifetime of the excited states that is assumed and $\varphi(t,\tau)$ the Light-Induced Phase (LIP) caused by the AC Stark effect~\cite{ott-2013,chini-prl:2012,chen-2013b}, which can be approximated for small IR intensities ($\Omega(t)\ll1$) as
\begin{equation}
\varphi(t,\tau) \approx \frac{1}{2\Delta}\int_\tau^tdt' \Omega^2(t').\label{eq:lip_approx}
\end{equation}

The dipole moment $d(t)$ has two contributions: (i) $C_n$ contains only terms that depend on the square of the transition dipole moments connecting the ground to either of the excited states, $d_{0n}^{2}$, and (ii) $N_n$ contains only terms that depend on the product of them, $d_{01}d_{02}$. For systems that obey the Laporte rule (i.e. a centrosymmetric system), $N_n$ vanishes since $d_{01}d_{02}d_{12}\equiv0$. This implies that either $d_{01}d_{02}=0$ or $d_{12}=0$, in which case a lack of coupling between the two excited-states prohibits an IR-induced state-mixing, i.e. $\sin\alpha(t)=\sin\alpha(\tau)=0$.

\section{Results and Discussion}~\label{sec:discussion}
\begin{figure*}[tbh]
    \includegraphics[width=\textwidth]{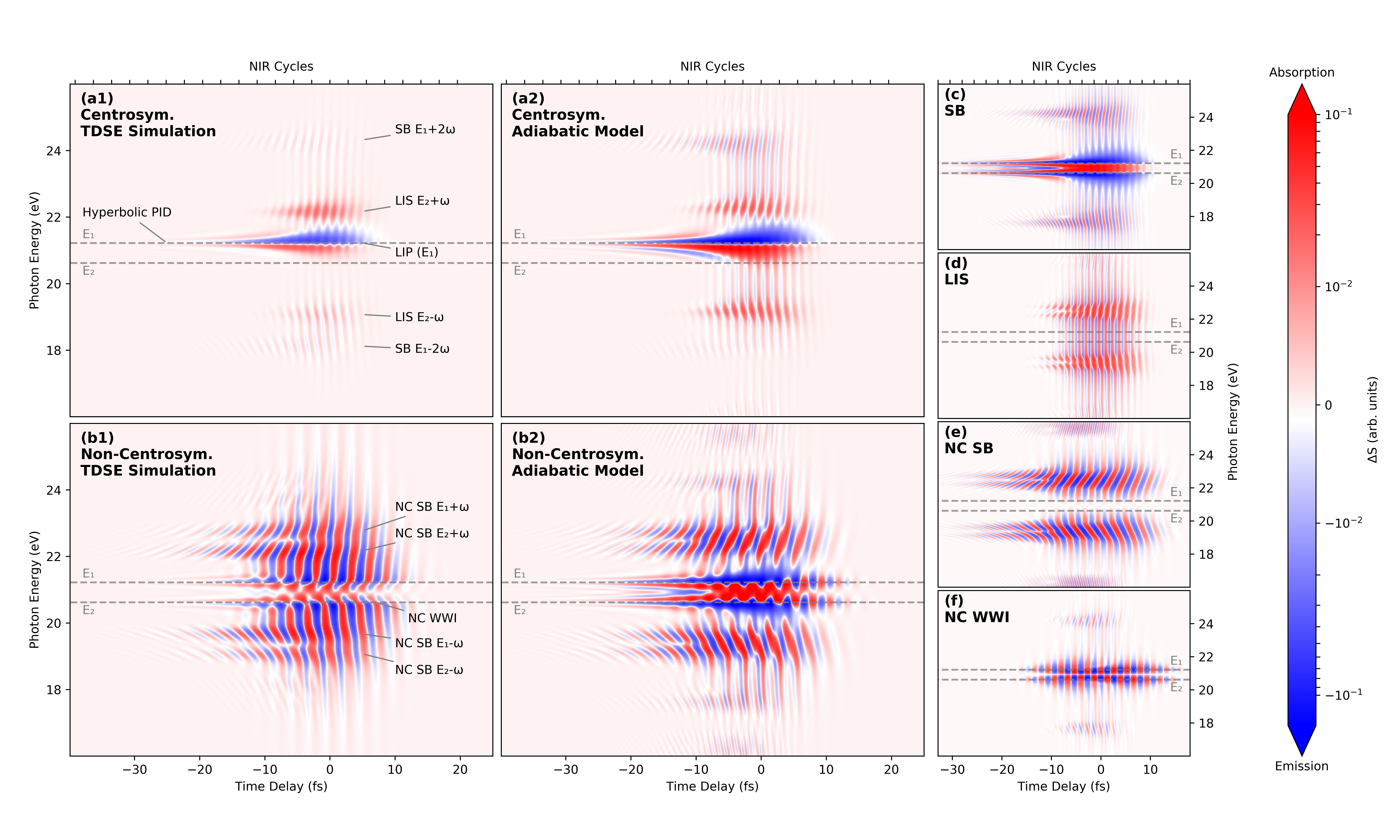}
    \caption{ATAS spectrograms for the centrosymmetric (a) and for the non-centrosymmetric case (b), for fixed CEP (0), obtained by the TDSE simulation (left column) and the adiabatic model (center column). In the right column (c-f), the individual contributions separated in the adiabatic model are shown. Positive time delays correspond to the IR pulse arriving before the XUV pulse.}\label{fig:1}
\end{figure*}
In figure\,\ref{fig:1}, alongside the results obtained by numerically solving the TDSE, ATAS spectrograms are shown in panels (a2),(b2) that were obtained using the analytical approach outlined in equations (\ref{eq:adiabatic_basis})-(\ref{eq:lip_approx}), by numerical solution of the LIP integral (Eq.~\eqref{eq:lip_approx}), by Fourier transformation of the dipole moment (Eq.~\eqref{eq:adia_dipole}) and by calculation of the change of spectral response. The analytical calculation qualitatively reproduces the results of the numerical TDSE solution, both for the centrosymmetric case ((a) panels) and the non-centrosymmetric case ((b) panels). 

The analytic solution lends itself to prying apart the different mechanisms that underlie the calculated ATAS spectrograms. Both $C_n$ and $N_n$ contain two terms each that depend on the IR-induced state-mixing at the time of excitation by the XUV pulse (described by $\alpha(\tau)$) and during the subsequent evolution (described by $\alpha(t)$). The effect of the mixing-angle terms on the transient absorption spectrum can be clarified by expanding $\alpha(t)$ in orders of $\Omega(t)$, the strength of the IR-induced coupling:
\begin{equation}
\begin{aligned}
\cos\alpha(t) &= 1-\frac{\Omega^2(t)}{8\Delta^2}+\mathcal{O}(\Omega^4)\\
\sin\alpha(t) &= \frac{\Omega(t)}{2\Delta}+\mathcal{O}(\Omega^3)
\end{aligned}
\label{eq:omega_series}
\end{equation}
Upon Fourier transform into the spectral domain, time-dependent terms proportional to $\Omega(t)$ and $\Omega^2(t)$ become sidebands, displaced by $\pm\omega_{\mathrm{IR}}$ and $\pm2\omega_{\mathrm{IR}}$. Since the femtosecond-duration IR pulse is inherently non-monochromatic, the spectral width of these sidebands will be given by a convolution of their pulse spectral envelope and the natural linewidth. The sidebands are modulated as a function of time-delay by the $\tau$-dependent terms.

The individual contributions from the four terms constituting $C_n$ and $N_n$ are separately shown in the right column of figure\,\ref{fig:1}. The centrosymmetric case ($d_{02} = 0$) was recently discussed by Rørstad \textit{et al.}~\cite{rorstad-2017} but is briefly covered here as well in order to distinguish its features from the additional characteristics that emerge in the non-centrosymmetric case. Panel\,(c) is obtained by exclusively considering the term containing $\cos\alpha(\tau)\cos\alpha(t)$ in $C_n$ (Eq.~\eqref{eq:Sn}). The leading term ($=1$) in the series expansion of $\cos\alpha(t)$ (Eq.~\eqref{eq:omega_series}) results in a strong LIP effect on the resonance absorption line. In the centrosymmetric case, this LIP results from the AC Stark shift of the bright $E_1$ state, caused by its coupling to the dark $E_2$ state. The LIP changes the interference between the free-induction decay and the incident XUV field, resulting in a delay-dependent (sub-cycle) reshaping of the absorption line, typically from Lorentzian (symmetric) to Fano-like (dispersive)~\cite{ott-2013}. This can be seen close to the energy of $E_1$ in figure\,\ref{fig:1}(a1) and (a2). Since both $E_1$ \textit{and} $E_2$ are bright states in our non-centrosymmetric model, their role is interchangeable and they both exhibit a LIP effect, as can be seen in Panel\,(c). At longer delays, the accumulated LIP is constant, but the fast oscillating phase from the term $e^{iE_n\tau}$ leads to a Perturbed Induction Decay (PID), manifested by hyperbolically converging absorption and emission features \cite{wu-2016}. The structures displaced by $\pm\,2\omega_{\mathrm{IR}}$ from the field-free resonance (cf. SB $E_1\pm\,2\omega$ in figure\,\ref{fig:1}(a1)) are the sidebands that originate from an excitation of two-photon dressed states \cite{chen-2013,rorstad-2017}. Panel\,(d) is obtained by exclusively considering the term containing $\sin\alpha(\tau)\sin\alpha(t)$ in $C_n$ and exhibits sidebands displaced by $\pm\,\omega_{\mathrm{IR}}$ with respect to the resonant energy. These result from two-photon XUV\,$\pm$\,IR excitation of the excited states, in which XUV absorption into virtual states at $E_n\pm\omega_{\mathrm{IR}}$ is accompanied by IR photon absorption/emission. In the centrosymmetric case these transitions only appear for the dark state (i.e. $E_2$, see figure\,\ref{fig:1}(a)). Their dependence on $\sin\alpha(\tau)$ (i.e. the fact that they require a non-zero IR field at the time of the XUV excitation) explains their appearance only at time overlap. Note that while one might expect an $1\omega_{\mathrm{IR}}$-oscillation due to the $\sin\alpha(\tau)$ term, the delay-dependent interference between the XUV field and the dipole moment leads to a hyperbolic term after Fourier transformation~\cite{wu-2016,rorstad-2017}, which combines to the observed $2\omega_{\mathrm{IR}}$ modulation (see Appendix~\ref{ax:hyperbole}). 

As seen from $C_n$ (Eq.~\eqref{eq:Sn}), the terms that contribute to the description of the spectrogram in the centrosymmetric case contain a product of sines and cosines of two mixing angles, one evaluated at time $t$, and one evaluated at delay $\tau$. In the adiabatic basis (Eq.~\eqref{eq:adiabatic_eigenstates}) these can be understood as (i) an IR-dependent projection from each of the field-free states onto their adiabatically evolving states ($e^{iE_n(t-\tau)} \cos\alpha(\tau)$) and then back to the field-free states after the interaction with the IR field ($\cos\alpha(t)$) and (ii) such a projection from each of the field-free states onto the other adiabatically evolving state ($e^{iE_m(t-\tau)} \sin\alpha(\tau)$) and then back again to the original field-free state ($\sin\alpha(t)$). In contrast, in the non-centrosymmetric case the additional term $N_n$ (Eq.~\eqref{eq:An}) describes a coherent XUV excitation of both states. The terms containing a product of sines and cosines, evaluated at time $t$ and delay $\tau$ can be interpreted in terms of a transfer from one field-free state to the other via a coherent superposition of intermediate adiabatic states.

A prominent additional feature introduced in the spectrograms for the non-centrosymmetric case is a pair of sidebands at energies lying one IR photon above and below the field-free resonance energies of both bright excited states ($E_n\pm\omega$ in figure\,\ref{fig:1}(b1), (b2)). These non-centrosymmetric SBs (NC SBs) are singled out in panel\,(e), obtained by exclusively considering the term containing $\cos\alpha(\tau)\sin\alpha(t)$ in $N_n$ (Eq.~\eqref{eq:An}). NC SBs result from an extension of the centrosymmetric SB mechanism: They appear at energies of $\pm\,\omega_{\mathrm{IR}}$ displaced from the excited states, in contrast to $\pm\,2\omega_\mathrm{IR}$ in the centrosymmetric case, and originate from the breakdown of the Laporte rule \cite{rorstad2018:pra}. Importantly, the NC SBs differ from the LIS found at the same energies in that they are observed also outside of temporal overlap, i.e. when the IR field arrives after XUV field. Since only a single IR photon is involved, the resulting modulation of the NC SB with $\tau$ occurs with the periodicity of the IR field. Note that while, from the expansion of $\cos\alpha(\tau)$, oscillations of the sidebands at the periodicity $2\omega_{\mathrm{IR}}$ might be expected, it is the hyperbolic interference condition with the XUV field that results in the $\omega_{\mathrm{IR}}$ oscillations.

The second prominent additional feature introduced in the spectrograms for the asymmetric case is a modification of the absorption strength directly at the field-free resonance energies (label `NC WWI' (non-centrosymmetric which-way interference) in figure\,\ref{fig:1}(b1)). The effect is also seen in panel\,(f), obtained within the adiabatic model by exclusively considering the term containing $\sin\alpha(\tau)\cos\alpha(t)$ in $N_n$ (Eq.~\eqref{eq:An}). Due to the constant in the expansion of $\cos\alpha(t)$ (Eq.~\eqref{eq:omega_series}) the modulation as a function of delay introduced by $\sin\alpha(\tau)$ remains spectrally at the field-free resonance energy. This modulation follows the IR field $\varepsilon_{\mathrm{IR}}(\tau)$ ($\omega_{\mathrm{IR}}$-oscillations), as seen from a series expansion of $\sin\alpha(\tau)$. NC WWI stems from the interference of two processes with which population is transferred to the same final state ($E_1$ or $E_2$): One-color (XUV-only) and two-color (XUV\,$\pm$\,IR) excitation, where the XUV absorption in the two-color pathway leads to a NC SB (see above). NC WWI does not exist for centrosymmetric systems due to the Laporte rule and has not been observed experimentally in transient absorption to date. Note also that the $\Omega^2(t)$ term in the expansion of $\cos\alpha(t)$ results in another set of very weak sidebands displaced by $2\omega_{\mathrm{IR}}$ from the resonance energy, seen in figure\,\ref{fig:1}(f).

\begin{figure}[tb]
    \includegraphics[width=1.0\columnwidth]{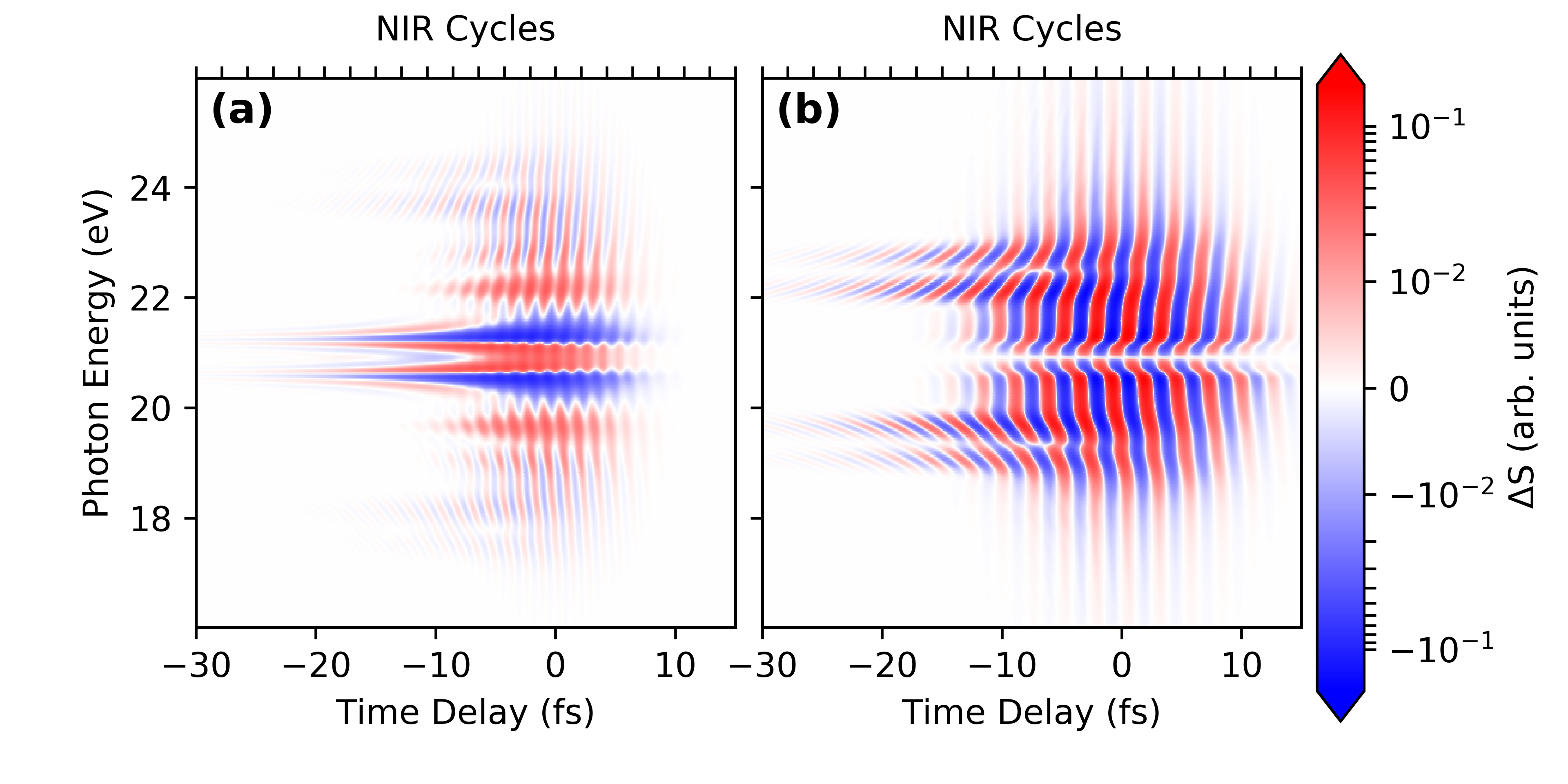}
    \caption{Numerically obtained ATAS spectograms, which are (a) the sum of two results derived for opposite CEP (0 and $\pi$) or, equivalently, the result of calculations for up- and down-oriented molecules at fixed CEP (0) and (b) the difference between these calculations.}\label{fig:3}
\end{figure}

Importantly, due to the WWI for non-centrosymmetric systems, the CEP of the IR electric field $\varepsilon_{\mathrm{IR}}(\tau)$ controls the delay-dependent modulation of the differential absorbance at the field-free resonance energies $E_1$ and $E_2$. Changing the CEP of the IR laser pulse by $\pi$ leads to inversion of the linear electric field ($\varepsilon_\textrm{IR}(t,\phi=0) = -\varepsilon_\textrm{IR}(t,\phi=\pi)$) and therefore of $\Omega(t)$. Terms depending on $\sin\alpha(t)$ and $\sin\alpha(\tau)$ will change sign when the sign of $\Omega$ changes (odd-terms), while the terms depending on $\cos\alpha(t)$ and $\cos\alpha(\tau)$ will remain unchanged under inversion of $\Omega$ (even-terms). In figure\,\ref{fig:3} the effect of controlling the CEP is explored in more detail. The spectrogram in panel (a) was obtained by incoherently adding the result of two numerical simulations with a CEP of the IR pulse of 0 and $\pi$. The observed spectrogram resembles the one calculated for the centrosymmetric case, with the distinction that both excited states are bright, as was the case for the individual contributions described by $C_n$ (figure\,\ref{fig:1}(c) and (d)). In figure\,\ref{fig:3}(b) we show the difference of two spectrograms obtained for a CEP of 0 and $\pi$. Since the even-terms that give rise to $2\omega_\textrm{IR}$ oscillations cancel out, only effects featuring $\omega_{\mathrm{IR}}$-oscillations remain, which result from the non-centrosymmetric term $N_n$, i.e. the contributions shown in figure\,\ref{fig:1}(e) and (f). This means that by comparing measurements with a controlled CEP, the signal depending on the non-centrosymmetric term $N_n$ can be isolated.

We emphasize that, as stated above, the results described were obtained for perfectly spatially oriented non-centrosymmetric molecules. When the orientation of the system is reversed, the sign of the transition dipole moment is inverted as it is antiparallel to the IR electric field, i.e. $\vec{d}_{nm}\cdot\vec\varepsilon_\textrm{IR}(t)=-d_{nm}\varepsilon_{IR}(t)$, thus leading to an inversion of $\Omega(t)$. This has the equal result to a CEP shift by $\pi$. Therefore, a spatial distribution of samples that is aligned (i.e. even distribution of molecules oriented parallel and antiparallel to the electric field) is equal to the spectogram shown in figure\,\ref{fig:3}(a) and figure\,\ref{fig:3}(b) also depicts the difference between the results for the two orientation directions. This shows that to experimentally observe the described features that rely on the non-centrosymmetric terms $N_n$, the centrosymmetry of the sample needs to be broken on a macroscopic level, i.e. the molecular sample needs to be oriented. Molecular alignment is not sufficient to allow for observations of the $\omega_{\mathrm{IR}}$-oscillations. However, as seen in figure\,\ref{fig:3}(a) even if the inversion symmetry is not broken in the laboratory frame, the coupling between bright states, which is possible due to a non-centrosymmetric molecular frame, leads to features that need to be considered in the analysis of molecular ATAS.

Due to this difference of the observability of odd- and even-terms, ATAS with CEP-stable and -controlled pulses on non-centrosymmetric molecules has potential as an all-optical diagnostic tool of the spatial orientation of molecules in a gas phase sample, introduced by a laser pulse in a non-adiabatic orientation scheme~\cite{ghafur-2009,holmegaard2010:natphys}. Moreover, we point out that in the case of an oriented sample (e.g. a laser-induced gas phase sample or a naturally oriented solid-state sample), ATAS becomes sensitive to CEP-stability, since for a uncontrolled CEP the $1\omega_\textrm{IR}$ signals in ATAS quickly average out in a measurement over multiple laser shots.

Finally, the strict symmetry argument depending on the lack of centrosymmetry creates an intriguing opportunity to study ultrafast changes in the electronic symmetry of molecules. Centrosymmetry-sensitive transient absorption could hereby be used as a probe to observe charge localization during dissociation~\cite{kling-2006} or photon-induced symmetry breaking~\cite{martin-2007} in molecules.

\newpage
\appendix
\section*{Acknowledgments}
We thank Dr. Oleg Kornilov (MBI Berlin) for help at early stages of the project and in addition with Dr. Serguei Patchkovskii (MBI Berlin) for stimulating discussion and for commenting on the manuscript.

\section{Parameters for Numerical Solution}\label{ax:tdse}

The XUV and IR electric fields used for the simulations were Gaussian envelope pulses:
\begin{equation}
\varepsilon(t,\tau) = \sqrt{I} e^{-(\frac{t-\tau}{T})^2} \cos[\omega (t-\tau)-\phi],\label{eq:laser_field}
\end{equation}
where $T_\textrm{XUV,IR} = \mathrm{FWHM}/2\sqrt{\ln2}$ defines the temporal width of the pulse envelope in terms of its full-width at half-maximum (FWHM), $I_\textrm{XUV,IR}$ is the pulse (peak) intensity and $\omega_\textrm{XUV,IR}$ the central frequency, respectively. For the IR pulse $\tau$ defines the XUV-IR time-delay, while the XUV pulse is fixed centered around $t=0$, i.e. $\tau=0$ corresponds to XUV-IR time overlap and $\tau>0$ ($<0$) refers to the situation where the IR pulse preceeds (trails) the XUV pulse.  The parameters for the two pulses are listed in table~\ref{tab:theory_laser}. The carrier-to-envelope phase $\phi$ was zero unless stated otherwise.

\begin{table}[htbp]

\begin{ruledtabular}
\begin{tabular}{lrr}
  & IR-Pulse & XUV-Pulse \\
Central Frequency & 1.55\,eV & 20.0\,eV \\
Pulse Duration (FWHM) & 7.0\,fs & 40\,as \\
Peak Field Intensity & $10^{12}$\,W/cm$^2$ & $10^{9}$\,W/cm$^2$ \\
\end{tabular}
\end{ruledtabular}
\caption{Laser parameters used in the simulations.}\label{tab:theory_laser}
\end{table}

\section{Analytical Solution with Adiabatic Basis}\label{cha:theory_adiabatic}
We will derive the analytical expression for the time-dependent dipole using the adiabatic basis, as described in Eq.~\eqref{eq:adiabatic_basis}.
After the initial excitation from the ground-state by the XUV pulse, the IR-induced dynamics only involves the two excited states $E_1$ and $E_2$, and we therefore consider the reduced two-level system:
\begin{equation}
H_r = \left[\begin{matrix} E_R+\Delta & \Omega(t) \\ \Omega(t) & E_R-\Delta\\ \end{matrix} \right],
\end{equation}
where $E_R = (E_1+E_2)/2$, $\Delta = (E_1-E_2)/2$ and $\Omega(t)=d_{12}\varepsilon_{IR}(t)$.

Upon diagonalization, the time-dependent eigenenergies for the excited states are:
\begin{equation}
\begin{aligned}
E_{+} =& E_R + \sqrt{\Delta^2 + \Omega^2(t)},\\
E_{-} =& E_R - \sqrt{\Delta^2 + \Omega^2(t)},
\end{aligned}
\end{equation}

with normalized adiabatic eigenstates:
\begin{widetext}
\begin{equation}
\begin{aligned}\vert\varphi_+(t)\rangle=&\frac{\Delta + \sqrt{\Delta^{2} + \Omega(t)^{2}}}{\sqrt{\Omega(t)^{2} + \left(\Delta + \sqrt{\Delta^{2} + \Omega(t)^{2}}\right)^{2}}}\vert\Phi_1\rangle+\frac{\Omega(t)}{\sqrt{\Omega(t)^{2} + \left(\Delta + \sqrt{\Delta^{2} + \Omega(t)^{2}}\right)^{2}}} \vert\Phi_2\rangle\\ 
\doteq&\cos\alpha(t)\vert\Phi_1\rangle+\sin\alpha(t)\vert\Phi_2\rangle,\\
\vert\varphi_-(t)\rangle=&\frac{-\Omega(t)}{\sqrt{\Omega(t)^{2} + \left(\Delta + \sqrt{\Delta^{2} + \Omega(t)^{2}}\right)^{2}}}\vert\Phi_1\rangle+\frac{\Delta + \sqrt{\Delta^{2} + \Omega(t)^{2}}}{\sqrt{\Omega(t)^{2} + \left(\Delta + \sqrt{\Delta^{2} + \Omega(t)^{2}}\right)^{2}}} \vert\Phi_2\rangle \\
\doteq&-\sin\alpha(t)\vert\Phi_1\rangle+\cos\alpha(t)\vert\Phi_2\rangle,\label{eq:adiabatic_eigenstates_ax}
\end{aligned}
\end{equation}
\end{widetext}
where $\Phi_n$ are the field-free excited states, and $\alpha(t)$ is a state mixing angle (see Eq.~\eqref{eq:mixing_angle}).

To find the state-coefficients $c_+(t)$ and $c_-(t)$ of the adiabatic basis, the wavefunction is expressed in the field-free basis:
\begin{equation}
\vert \Psi(t) \rangle = b_0(t)\vert \Phi_0 \rangle + b_1(t)\vert\Phi_{1}\rangle + b_2(t) \vert\Phi_{2}\rangle.\label{eq:field-free-basis}
\end{equation}

Combining Eq.~\eqref{eq:adiabatic_basis} and Eq.~\eqref{eq:field-free-basis} with Eq.~\eqref{eq:adiabatic_eigenstates_ax} results in:
\begin{equation}
\begin{aligned}
b_1(t) &= c_+(t)e^{i\theta_{+}}\cos\alpha(t)
    - c_-(t)e^{i\theta_{-}}\sin\alpha(t),\\
b_2(t) &= c_+(t)e^{i\theta_{+}}\sin\alpha(t)
    + c_-(t)e^{i\theta_{-}}\cos\alpha(t).
\end{aligned}
\end{equation}
To obtain the temporal dependence of the excited state coefficients, the expression for the field-free wavefunction (Eq.~\eqref{eq:field-free-basis}) is inserted into the TDSE, leading to:
\begin{equation}
\begin{aligned}
i\dot{b_1}=& b_0 d_{01} \varepsilon_\textrm{XUV} + b_2 d_{12}\varepsilon_{IR}\\
i\dot{b_2}=& b_0 d_{02} \varepsilon_\textrm{XUV} + b_1 d_{12}\varepsilon_{IR}.\label{eq:adia_ode}
\end{aligned}
\end{equation}

A perturbative approach is used to solve for the coefficients~\cite{griffiths-1995}: If the system remains unperturbed, the entire population remains in the ground state: $b_0^{(0)}(t) = 1,  b_1^{(0)}(t) = b_2^{(0)}(t) = 0$ (zeroth-order approximation). To obtain the coefficients in first order, the zeroth-order wavefunction amplitudes are inserted into Eq.~\eqref{eq:adia_ode} and the set of equations is converted to a set of equations for the amplitudes in the adiabatic basis:
\begin{equation}
\begin{aligned}
\dot{c}_+(t)e^{i\theta_{+}(t)}\cos\alpha(t) - \dot{c}_-(t)e^{i\theta_{-}(t)}\sin\alpha(t) &= -id_{01}\varepsilon_\textrm{XUV},\\
\dot{c}_+(t)e^{i\theta_{+}(t)}\cos\alpha(t) + \dot{c}_-(t)e^{i\theta_{-}(t)}\sin\alpha(t) &= -id_{02}\varepsilon_\textrm{XUV}.
\end{aligned}
\end{equation}

Solving for $\dot{c}_\pm^{(1)}(t)$ gives:
\begin{equation}
\begin{aligned}
\dot{c}_+^{(1)}(t) &= -i\varepsilon_\textrm{XUV}e^{- i \theta_{+}(t)}\left(d_{01}\cos\alpha(t)+d_{02}\sin\alpha(t)\right),\\
\dot{c}_-^{(1)}(t) &= +i\varepsilon_\textrm{XUV}e^{- i \theta_{-}(t)}\left(d_{01}\sin\alpha(t)-d_{02}\cos\alpha(t)\right).
\end{aligned}
\end{equation}

Assuming that the initial excitation of the XUV pulses is sufficiently short with respect to the IR-induced dynamics, it can be approximated by a Dirac $\delta$-function~\cite{wu-2016,rorstad-2017}: $\varepsilon_\textrm{XUV}(t) \to I_0\delta(t-\tau)$, and
$c_\pm^{(1)}(t)$ can be obtained by direct integration:
\begin{equation}
\begin{split}
c_+^{(1)}(t) &= \int_{\tau}^t \dot{c}_+^{(1)}(t')\,dt'\\ &= -i\vartheta(t-\tau) \left(d_{01}\cos\alpha(\tau)+d_{02}\sin\alpha(\tau)\right),\\
c_-^{(1)}(t) &= \int_{\tau}^t \dot{c}_-^{(1)}(t')\,dt'\\ &= i\vartheta(t-\tau) \left(d_{01}\sin\alpha(\tau)-d_{02}\cos\alpha(\tau)\right),\label{eq:adiabatic_coeff}
\end{split}
\end{equation}
where $\vartheta(t)$ is the Heaviside function. Note that in these expressions the state-mixing angle, which defines the relation between the adiabatic eigenstates and the field-free eigenstates, is no longer a function of $t$, but a function of the XUV-IR delay $\tau$.

Ultimately, we are interested in the time-dependent dipole expressed in terms of the field-free basis, which can be written as:

\begin{equation}
\begin{split}
\langle d(t) \rangle = & d_{01}b_1(t)+d_{02}b_2(t) + c.c. \\ = & d_{01}[c_+(t)e^{i\theta_{+}}\cos\alpha(t)
    - c_-(t)e^{i\theta_{-}}\sin\alpha(t)]\\ + & d_{02}[c_+(t)e^{i\theta_{+}}\sin\alpha(t)
    + c_-(t)e^{i\theta_{-}}\cos\alpha(t)] + c.c.
\end{split}\label{eq:dipole_fieldfree}
\end{equation}

In the presented derivation, the time dependence of the dynamic phase term has so far been ignored. Inserting the time-dependent energy of the adiabatic states into Eq.~\eqref{eq:adiabatic_basis} allows to rewrite the state-dependent exponential term as:

\begin{equation}
\begin{split}
e^{i\theta_\pm} + c.c. = & e^{-i\int_\tau^t  E_0 \pm \sqrt{\Delta^2 + \Omega^2(t')}\,dt'} + c.c.\\
\approx & e^{-i(E_0\pm\Delta)(t-\tau)}e^{\mp i\varphi(t,\tau)}+c.c,\label{eq:dyn_phase_to_lip}
\end{split}
\end{equation}
where the fast-oscillating terms $e^{-i(E_0\pm\Delta)(t-\tau)}+c.c.$, i.e. the field-free resonant energy of the excited states, lead to a resonant frequency response at $-E_0\pm\Delta$ and $E_0\mp\Delta$ in the complex conjugated terms. $\varphi(t,\tau)$ is the so called Light-Induced Phase (LIP) caused by the AC Stark effect, which can be approximated for small IR intensities ($\Omega(t)\ll1$, see Eq.~\eqref{eq:lip_approx}). To account for the decay of the excited states, an imaginary energy $i\Gamma/2$ is added to the energy eigenvalues.

Inserting Eq.~\eqref{eq:dyn_phase_to_lip} with the added imaginary energy term and Eq.~\eqref{eq:adiabatic_coeff} into Eq.~\eqref{eq:dipole_fieldfree}, we obtain the full time-dependent dipole, as described by Eq.~\eqref{eq:adia_dipole}. As noted in the main text, numerical methods were used to solve the LIP integral and to calculate the Fourier transform of the time-dependent dipole.

\section{Delay-Dependent Interference with XUV-Field}\label{ax:hyperbole}
The spectral representation of the time-dependent dipole and the spectral response can be calculated analytically by further approximations. To investigate the delay-dependentent spectral response due to the interference with the XUV field, we will demonstrate this for the light induced structures (LIS).

As identified in the main text, the LIS originates from the $\sin\alpha(t)\sin\alpha(\tau)$-terms of $C_n$ in Eq.~\eqref{eq:adia_dipole}, i.e. the part of the time-dependent dipole:
\begin{equation}
\begin{split}
d(t)_\textrm{LIS}=&-i\theta(t-\tau)e^{-\frac\Gamma2 (t-\tau)}e^{i(E_m)(t-\tau)}e^{(-1)^n i\varphi}\\
& d_{0n} \sin\alpha(t)\sin\alpha(\tau)+\textrm{c.c.}
\end{split}
\end{equation}

If the LIP $\varphi$ is assumed to be a constant function of time, the first part of the time-dependent dipole can be transformed into the spectral domain:
\begin{equation}
\begin{split}
\mathcal F&[i\theta(t-\tau)e^{-\frac\Gamma2 (t-\tau)}e^{i(E_m)(t-\tau)}e^{(-1)^n i\varphi} ]\\
 &= e^{-i\tau E_m}\frac{i e^{(-1)^n i\varphi}}{\sqrt{2\pi}(\frac\Gamma 2+i(\omega-E_m)}.\label{eq:lip_fourier}
 \end{split}
 \end{equation}
Note that we have dropped the complex conjugated terms, since they correspond to identical features at negative frequencies due to the Hermitian property of the Fourier transformation.

For the second part, the IR electric field is approximated as monochromatic, so that the Rabi-frequency can be described as:
\begin{equation}
\Omega(t) = d_{12}E_0\cos(\omega_\textrm{IR}t)
\end{equation}
where $E_0$ is the field amplitude. We express the mixing-angles using the series expansion (Eq.~\eqref{eq:omega_series}):
\begin{equation}
\begin{split}
\sin\alpha(\tau)\sin\alpha(t)&\approx \frac{1}{4\Delta^2}\Omega(\tau)\Omega(t)\\
&= \frac{1}{4\Delta^2}d_{12}^2E_0^2 \cos(\omega_\textrm{IR}\tau)\cos(\omega_\textrm{IR}t).
\end{split}
\end{equation}

As discussed in the main text, the time-dependent mixing-angle terms will lead to the generation of sidebands after Fourier-transformation, while the delay-dependent terms remain:
\begin{equation}
\begin{split}
\mathcal F&\left[\frac{1}{4\Delta^2}d_{12}^2E_0^2 \cos(\omega_\textrm{IR}\tau)\cos(\omega_\textrm{IR}t)\right](\omega)\\
&= \frac{1}{4\Delta^2}\cos(\omega_\textrm{IR}\tau)\sqrt{2\pi}d_{12}^2E_0^2\frac{\delta(\omega-\omega_\textrm{IR})+\delta(\omega+\omega_\textrm{IR})}{2}.\label{eq:sidebands_appendix}
\end{split}
\end{equation}

Both parts of the spectral dipole are combined using the convolution theorem:
\begin{equation}
\mathcal F[f\cdot g](\omega) = \mathcal F[f](\omega)*\mathcal F[g](\omega),
\end{equation}
where:
\begin{equation}
(f*g)(x) = \int_{-\infty}^\infty f(y)g(x-y)\,dy.
\end{equation}

The spectral dipole of the LIS then is:
\begin{widetext}
\begin{equation}
\begin{aligned}
d(\omega)&_{\mathrm{LIS}} \approx
-\mathcal F[i\vartheta(t-\tau)e^{-\frac\Gamma2 (t-\tau)}e^{iE_m(t-\tau)}e^{(-1)^ni\varphi}]* d_{02}^2\mathcal F\left[\frac{1}{4\Delta^2}d_{12}^2E_0^2 \cos(\omega_\textrm{IR}\tau)\cos(\omega_\textrm{IR}t)\right]\\
=& -\left[e^{-iE_m\tau}\frac{i e^{(-1)^ni\varphi}}{\sqrt{2\pi}(\frac\Gamma 2+i(\omega-E_m))}\right]*\left[\frac{\sqrt{2\pi}d^2_{0n}d^2_{12}E^2_0}{4\Delta^2}\cos(\omega_\textrm{IR}\tau)\frac{\delta(\omega-\omega_\textrm{IR})+\delta(\omega+\omega_\textrm{IR})}{2}\right]\\
=& -i\frac{d^2_{0n}d^2_{12}E^2_0}{8\Delta^2}\cos(\omega_\textrm{IR}\tau) e^{-iE_m\tau}e^{(-1)^ni\varphi}\frac{1}{(\frac\Gamma 2+i(\omega-E_m\pm\omega_\textrm{IR}))},
\end{aligned}
\end{equation}
\end{widetext}
where we have used the translational properties of the Dirac-delta function in convolutions. 

However, the observable signal is:
\begin{equation}
    S(\omega,\tau) = \operatorname{Im}\left[\frac{d(\omega)}{\varepsilon_\textrm{XUV}(\omega)}\right].
\end{equation}
Since a Dirac-delta pulse was assumed in the deviation of the adiabatic solution $\varepsilon_\textrm{XUV}(\omega)\propto e^{-i\omega\tau}$.
The additional fast oscillating term $e^{i\omega\tau}$ will therefore need to be considered.
This modifies the observable delay-dependent dynamic at the light-induced structures and sidebands. For the (centrosymmetric) LIS, the temporal behavior at the sideband energy is:
\begin{equation}
\begin{split}
S(E_m\pm\omega_\textrm{IR},\tau)&\propto \Im \left[\frac{i\cos(\omega_\textrm{IR}\tau)e^{-iE_m\tau}}{e^{-i(E_m\pm\omega_\textrm{IR})\tau}}\right]\\
&= \Re[\cos(\omega_\textrm{IR}\tau)e^{\pm i\omega_\textrm{IR}\tau}]\\
&= \pm\cos(\omega_\textrm{IR}\tau)^2.
\end{split}
\end{equation}

Therefore, at resonance, the symmetric light-induced structure will be modulated by $\cos(\omega_\textrm{IR}\tau)^2$.

\bibliography{main}

\end{document}